\documentclass[twocolumn,preprintnumbers,superscriptaddress,prl,amsmath,amssymb,showkeys]{revtex4}

\usepackage{graphicx}
\usepackage{dcolumn}
\usepackage{bm}
\usepackage{titlesec}

\begin{document}

\title{Flux growth utilizing the reaction between flux and crucible}

\author{J.-Q. Yan}
\affiliation{Materials Science and Technology Division, Oak Ridge National Laboratory, Oak Ridge, Tennessee 37831, USA}
\affiliation{Department of Materials Science and Engineering, University of Tennessee, Knoxville, Tennessee 37996, USA}

\date{\today}

\begin{abstract}
Flux growth involves dissolving the components of the target compound in an appropriate flux at high temperatures and then crystallizing under supersaturation controlled by cooling or evaporating the flux. A refractory crucible is generally used to contain the high temperature melt. The reaction between the melt and crucible materials can modify the composition of the melt, which typically results in growth failure, or contaminates the crystals. Thus one principle in designing a flux growth is to select suitable flux and crucible materials thus to avoid any reaction between them. In this paper, we review two cases of flux growth in which the reaction between flux and Al$_2$O$_3$ crucible tunes the oxygen content in the melt and helps the crystallization of desired compositions. For the case of La$_5$Pb$_3$O, Al$_2$O$_3$ crucible oxidizes La to form a passivating La$_2$O$_3$ layer which not only prevents further oxidization of La  in the melt but also provides [O] to the melt. For the case of La$_{0.4}$Na$_{0.6}$Fe$_2$As$_2$, it is believed that the Al$_2$O$_3$ crucible reacts with NaAsO$_2$ and the reaction consumes oxygen in the melt thus maintaining an oxygen-free environment.
\end{abstract}


\maketitle

\section{Introduction}

Materials synthesis and crystal growth in molten fluxes are compelling routes to novel materials with intriguing physical properties important for both scientific research and technological applications. Flux growth is also known as high temperature solution growth; the flux melts at high temperatures and acts as the solvent in a solution growth. In a typical flux growth, first the components of the target compound are dissolved in a flux, which is usually molten salt, oxide, or metal inside of a refractory crucible. Then, crystal growth takes place under supersaturation controlled by cooling or evaporating the flux. By dissolving the components of a desired compound in a flux with a low melting temperature, sometimes in controlled atmosphere or sealed ampoule, flux growth is good for the growth of a large variety of materials, such as compounds that would sublimate, compounds with a high melting temperature, volatile components, poisonous components, or a phase transition below the melting temperature.

Flux growth of oxides has been well developed in the last century. A good text book about this topic is "Crystal Growth from High-Temperature Solutions" by D. Elwell and H. J. Scheel.\cite{SolutionBook} There are also some good review articles. \cite{BWanklyn,ZurLoye,Fisher2012} Crystal growth and materials exploration from metallic flux were mainly developed in recent years with a major contribution from P. C. Canfield and Z. Fisk.\cite{Fisher2012, PaulandFisk, Fisk, decanting, DTA, sulphide, PaulandFisher,Chan,Jeitschko} Compared to the flux growth of oxides which is normally performed in air or flowing gas (O$_2$, N$_2$, or Ar), growth of intermetallic compounds from metallic flux or salt is more challenging because most intermetallics tend to oxidize in air and some are moisture sensitive, volatile, and/or poisonous.

In many cases, more than one kind of flux can mediate the crystal growth of one desired compound. Crystals of the same nominal composition but grown out of different fluxes might exhibit different physical properties, which could lead to confusions in understanding the intrinsic properties. One good example is the physical properties of K-doped BaFe$_2$As$_2$, one important member of FeAs-based superconductors. Kim et al\cite{BaKFe2As2} studied the low temperature specific heat of Ba$_{0.6}$K$_{0.4}$Fe$_2$As$_2$ single crystals grown out of Sn, In, and FeAs fluxes, and found that the physical properties depend on the preparation method. This illustrates the importance of growing crystals out of an appropriate flux. There are some important factors that have to be considered in selecting the right flux.\cite{SolutionBook, Fisk, Handbook} One of them is that the reaction between flux and crucible should be negligible.

In a flux growth, a refractory crucible is needed to hold the melt at high temperatures. Depending on the composition of the desired compound, the properties of flux, and the temperature profile, crucible materials can be metals (Fe, Ni, Ta, W, Mo, Pt), quartz, graphite, nitrides (BN), or oxides (Al$_2$O$_3$, MgO, ZrO$_2$). At high temperatures, a serious reaction between flux and crucible can destroy the crucible and ruin the growth. A minor reaction can modify the composition of the melt, which leads to the failure of the growth, or, in  most cases, contaminates the crystals as a substitution or an inclusion. Crystal contamination from crucible materials can significantly affect the physical properties. This is well illustrated by the growth of YBCO superconductors in an Al$_2$O$_3$ crucible. The BaO-CuO flux attacks Al$_2$O$_3$ crucibles and Al$_2$O$_3$ dissolved into the flux contaminates the crystals as substituent that suppresses  the superconducting transition temperature by over 20\,K.\cite{YBCO} Thus, when designing a flux growth, a good combination of flux and crucible is normally selected so that no or negligible reaction between flux and crucible is expected at high temperatures.

In this paper, we provide a different view of the reaction between flux and refractory crucible materials: in some special cases, the reaction can be employed to facilitate the growth of desired compositions. Specifically, we review the flux growth of La$_5$Pb$_3$O and La$_{0.4}$Na$_{0.6}$Fe$_2$As$_2$ in which the reaction between flux and Al$_2$O$_3$ crucible helps control the oxygen content in the melt for the growth of desired crystals.

\section{Crystal Growth of La$_5$Pb$_3$O}

La$_5$Pb$_3$O single crystals were grown out of a mixture of La, Co, and Pb. Starting materials La (99.999\% Ames Laboratory), Co (99.999\%, Alfa Aesar), and Pb (99.9999\%, Alfa Aesar) were placed in a 2 ml Al$_2$O$_3$ crucible with a molar ratio of 7:2:1. A catch crucible containing quartz wool was mounted on top of growth crucible and both were sealed in a silica ampoule under approximately 1/3 atmosphere of high pure argon gas. The sealed ampoule was heated to 1150$^{\circ}$C, stayed at 1150$^{\circ}$C for 4 hours, and then cooled to 850$^{\circ}$C over 60 hours. Once the furnace reached 850$^{\circ}$C, the excess flux was decanted from the rectangular crystals with the typical dimension of 0.5$\times$0.5$\times$2-4\,mm$^3$. Room temperature x-ray powder diffraction of pulverized single crystals confirmed single phase of as-grown crystals. The stoichiometry of as-grown crystals was confirmed with elementary analysis as well as single crystal x-ray and neutron diffraction performed at room temperature. Detailed crystal characterization and physical properties will be reported elsewhere.

\begin{figure}
\centering \includegraphics[width = 0.48\textwidth] {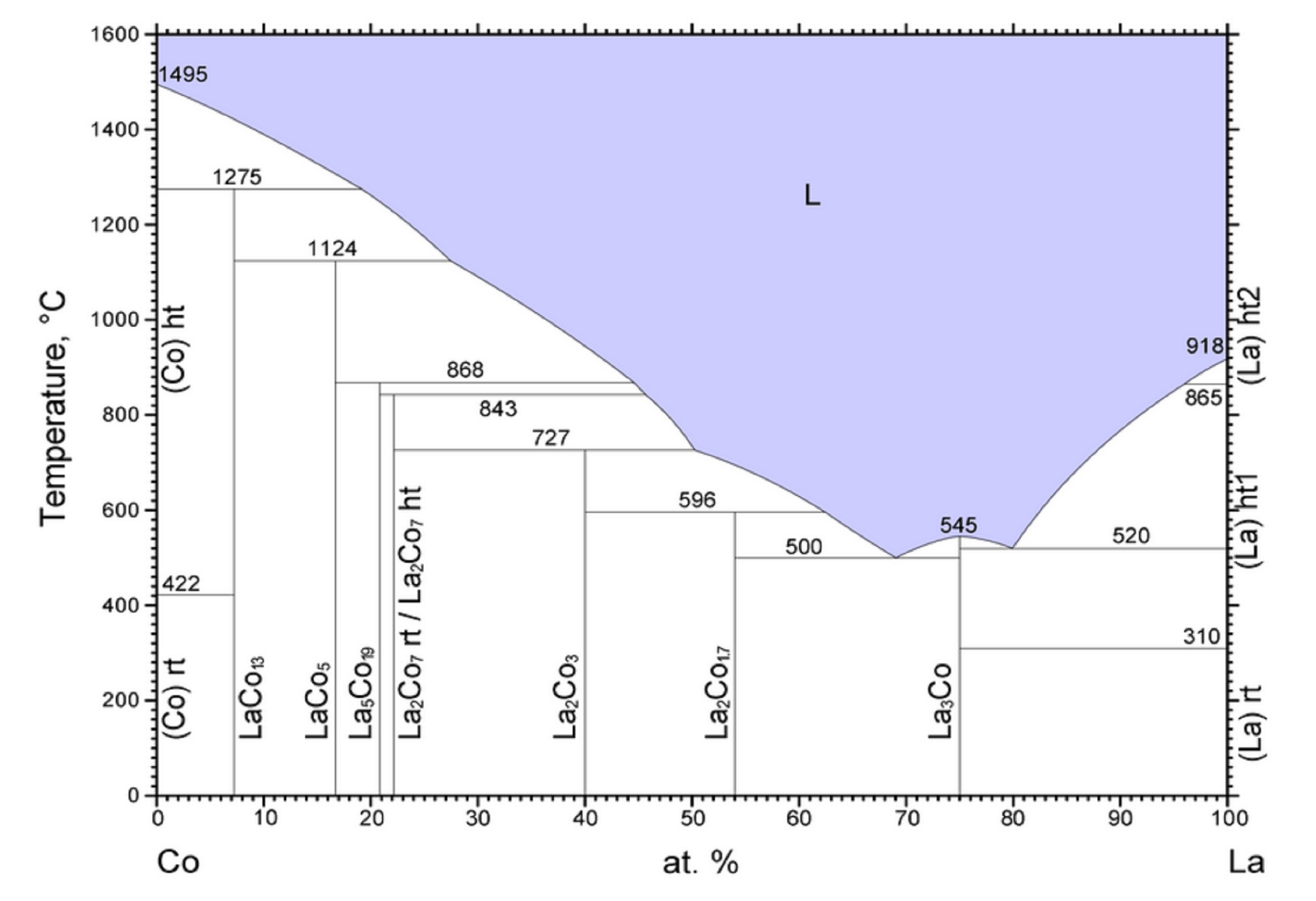}
\caption{(color online) La-Co phase diagram from ASM Alloy Phase Diagram Database.\cite{LaCo}} \label{PD-1}
\end{figure}

\begin{figure}
\centering \includegraphics[width = 0.48\textwidth] {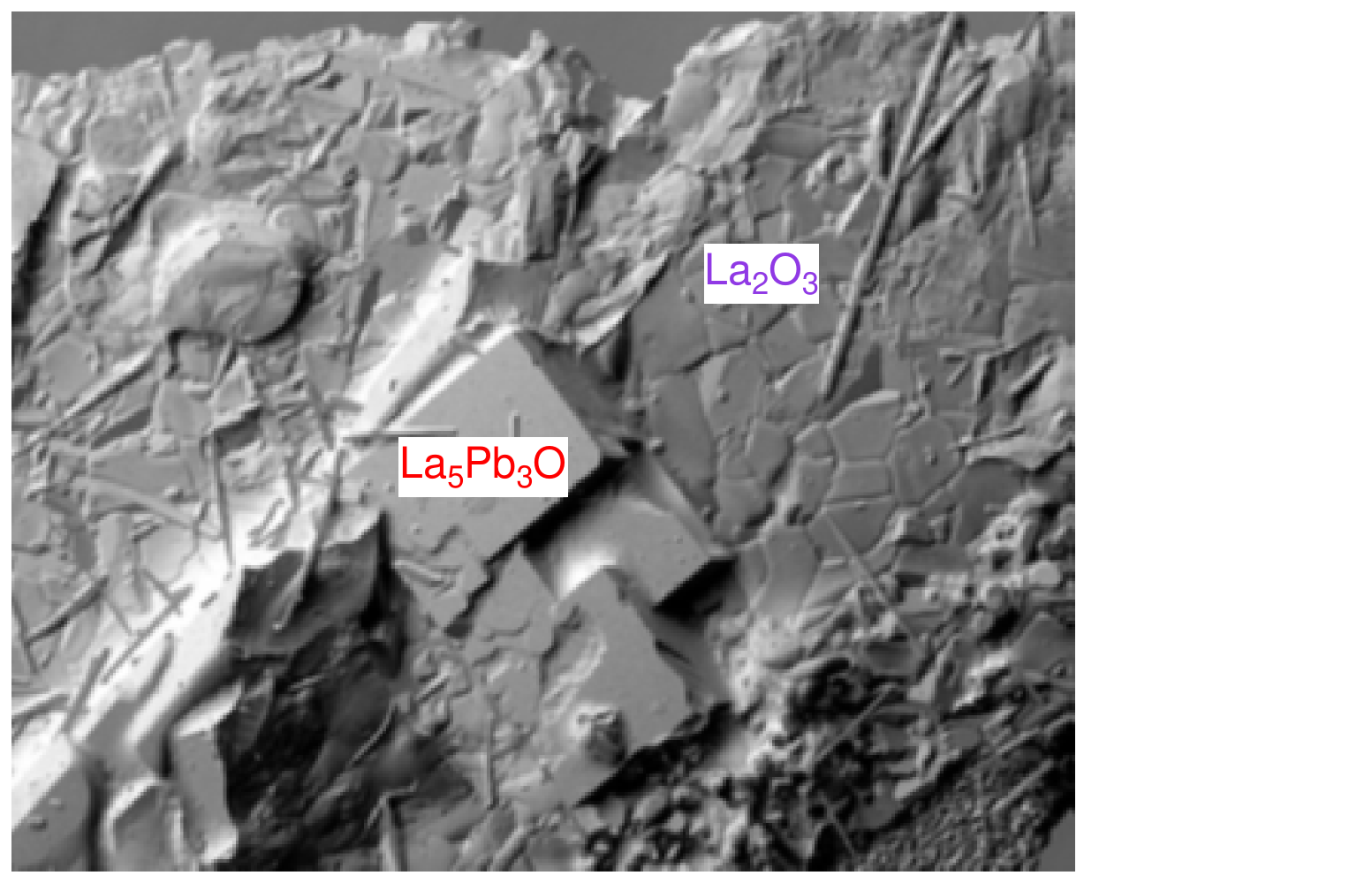}
\caption{(color online) SEM picture of La$_2$O$_3$ layer. La$_5$Pb$_3$O nucleates on La$_2$O$_3$ layer.} \label{La2O3-1}
\end{figure}

A similar procedure was previously employed to grow Ce$_5$Pb$_3$O single crystals.\cite{Ce5Pb3O} Once Co is absent in the starting materials, no Ce$_5$Pb$_3$O crystals could be obtained. This, together with the fact that no Co contamination was found in the as-grown crystals, suggest that Co is an important component in forming a proper flux mediating the growth of Ce$_5$Pb$_3$O. The successful growth of Co-free La$_5$Pb$_3$O in our study suggests a similar role of Co in the growth. Investigation of the La-Co and La-Pb binary phase diagrams suggests that Co is important in lowering the melting temperature of the mixture.  Figure\,\ref{PD-1} shows the La-Co binary phase diagram.\cite{LaCo} In a wide composition range 10$<$x$_{Co}$$<$55$\%$at, La$_{1-x}$Co$_x$ alloy melts below 850$^o$C. Even though Pb melts at a low temperature of ~320$^\circ$C, the melting temperature of La-Pb alloys increases rapidly when introducing La in to Pb melt; the line compound La$_5$Pb$_3$ melts congruently at 1450$^\circ$C.

It is interesting to note that no oxygen was introduced in the starting materials but the resulting crystals contain oxygen. After decanting, a thin layer of La$_2$O$_3$ was always observed inside of the alumina growth crucible and it takes the shape of the crucible. Figure\,\ref{La2O3-1} shows a SEM picture of part of the La$_2$O$_3$ layer. The high pure starting materials are not expected to introduce that much oxygen. The growth was performed in a sealed quartz tube with the highest homogenizing temperature of 1150$^o$C which is below the softening temperature of quartz. So it is unlikely that significant amount of oxygen diffuses into the quartz tube during crystal growth. As shown in Figure\,\ref{La2O3-1}, some quartz wool in the catch crucible fall down into the melt, which might provide oxygen to oxidize La. However, in a growth performed without using quartz wool inside of the catch crucible, both the La$_2$O$_3$ layer and La$_5$Pb$_3$O single crystals were obtained. Therefore, the alumina growth crucible, which has direct contact with the melt, is expected to be the only source for oxygen in La$_5$Pb$_3$O crystals and the La$_2$O$_3$ layer. As a rule of thumb, rare earth concentrations over 12\% in a wide variety of intermetallic fluxes react with Al$_2$O$_3$.\cite{12P} In the growth of La$_5$Pb$_3$O, there are 70\%at La in the melt and oxidization of La metal in the melt by the growth crucible is expected. This oxidization leads to the formation of La$_2$O$_3$ layer and also explains why the La$_2$O$_3$ layer takes the shape of the growth crucible.

\begin{figure}
\centering \includegraphics[width = 0.40\textwidth] {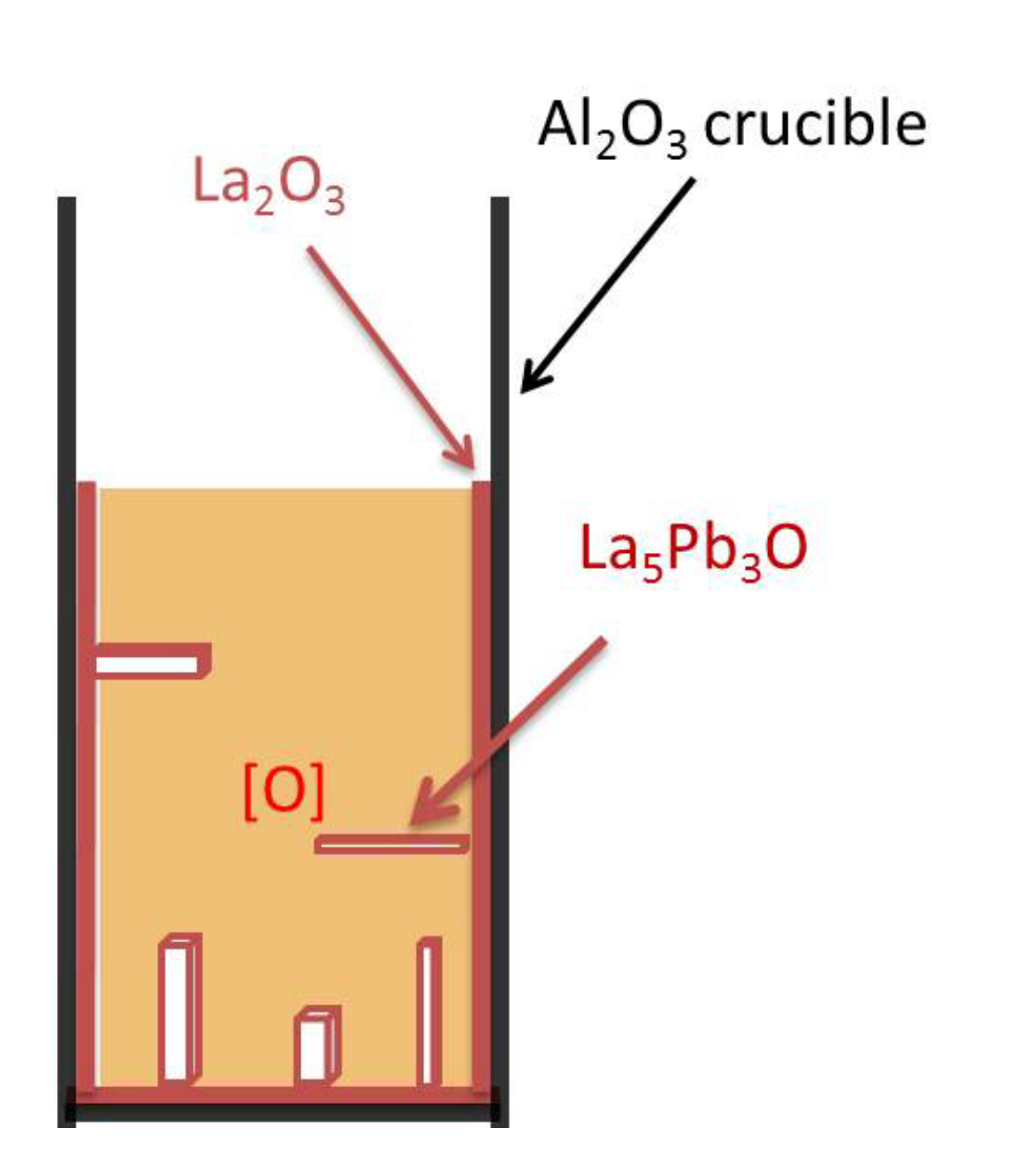}
\caption{(color online) Schematic picture illustrating the La$_2$O$_3$ layer formed at high temperatures.} \label{oxidization-1}
\end{figure}

\begin{figure}
\centering \includegraphics[width = 0.48\textwidth] {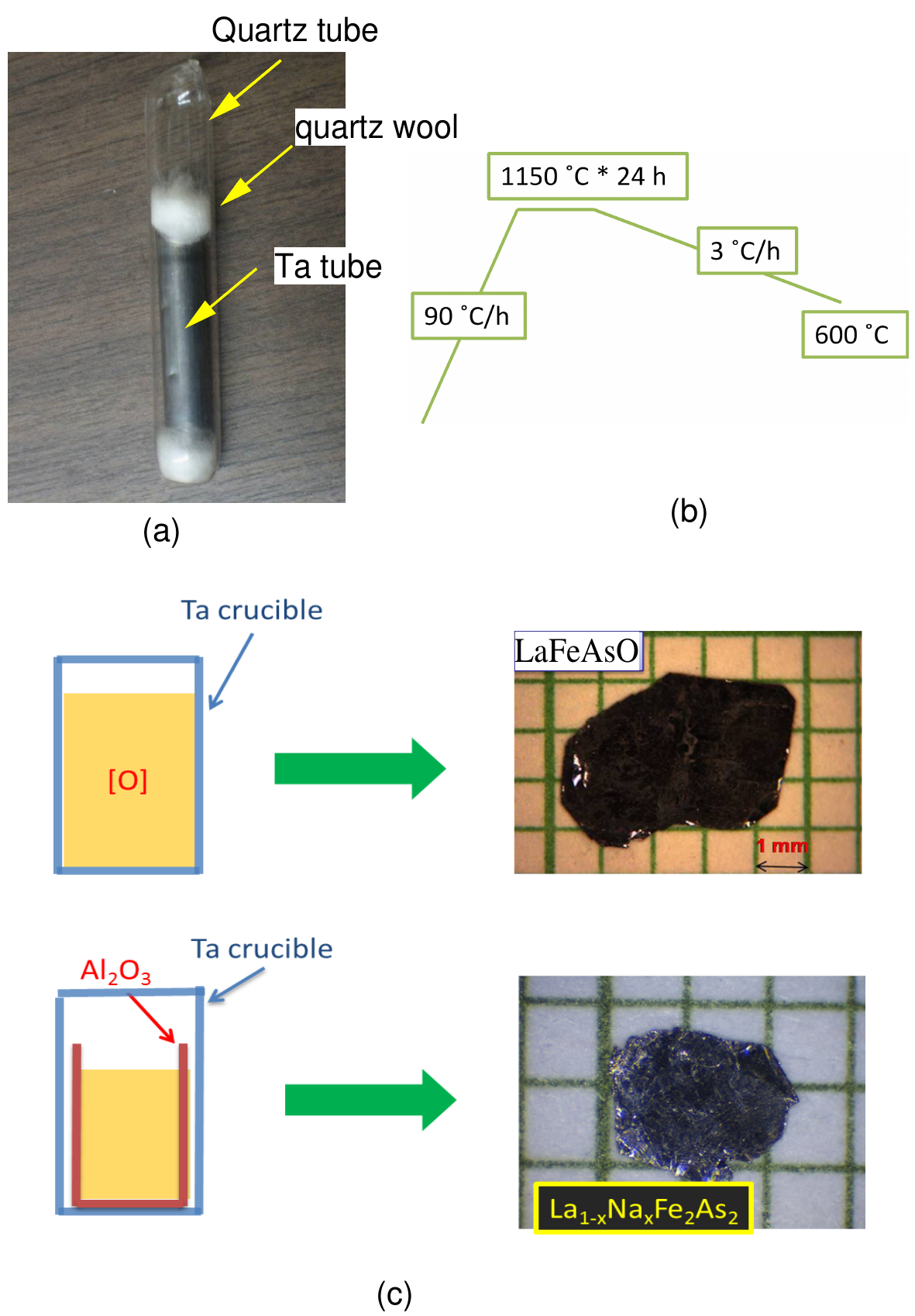}
\caption{(color online) (a) Growth ampoule. For the growth of LaFeAsO, all starting materials were loaded in the Ta tube. For the growth of La$_{0.4}$Na$_{0.6}$Fe$_2$As$_2$, the starting materials were held in an Al$_2$O$_3$ crucible inside of Ta tube.  (b) Temperature profile for both growths, (c) Schematic pictures illustrating the growth of LaFeAsO in Ta crucible and La$_{0.4}$Na$_{0.6}$Fe$_2$As$_2$ in Al$_2$O$_3$ crucible. Crystal picture of LaFeAsO is from ref[20] and that of La$_{0.4}$Na$_{0.6}$Fe$_2$As$_2$ from ref[19].} \label{LaNaFeAs-1}
\end{figure}

As shown in Figure\,\ref{La2O3-1}, the rectangular La$_5$Pb$_3$O grains nucleate on the surface of La$_2$O$_3$ layer. We believe that the oxidization layer is critical for the growth of La$_5$Pb$_3$O crystals: (1) it serves as the source of oxygen to the melt and controls the oxygen content in the melt by the reaction 2[La]\,+\,3[O]\,=\,La$_2$O$_3$(s); (2) it also serves as the passivating layer to prevent further oxidization of rare earth metal. The La/Co ratio in the starting materials stays close to the eutectic with the melting temperature about 640$^o$C (see Figure 1). Oxidization of La metal will modify the composition of the melt. The melting temperature of the melt increases when the flux becomes more Co-rich, and in the worst case La-Co binary compounds might precipitate. A decanting temperature of 850$^o$C was selected considering the composition change of the melt due to the oxidization of La during the growth. The lowest decanting temperature used in our growths is 825$^o$C. The successful decanting at 825$^o$C suggests that La:Co ratio should be less than 55$\%$at without considering the effect of alloying with Pb on the melting temperature of La-Co binary alloy.

Figure\,\ref{oxidization-1} illustrates how the reaction between growth crucible and flux helps the growth of La$_5$Pb$_3$O: at high temperatures, Al$_2$O$_3$ oxidizes La metal forming La$_2$O$_3$ passivating layer which controls the oxygen content in the melt; La$_5$Pb$_3$O crystals nucleate on the surface of La$_2$O$_3$ passivating layer. With a similar procedure, we have been able to grow \emph{R}$_5$Pb$_3$O (\emph{R}\,=\,La, Ce, Pr, Nd, and Sm) crystals. However, we failed to grow \emph{R}$_5$Pb$_3$O crystals with heavier rare earth ions even though a rare earth oxide layer also forms. This might signal limited oxygen solubility and/or diffusivity in \emph{R}-Co-Pb melt with heavy rare earth ions.

\section{Crystal Growth of La$_{0.4}$Na$_{0.6}$Fe$_2$As$_2$}

The growth and physical properties of La$_{0.4}$Na$_{0.6}$Fe$_2$As$_2$ single crystals were reported elsewhere.\cite{LaNaFe2As2} When looking for proper crucible materials for the growth of LaFeAsO superconductors out of NaAs flux, it was noticed that  La$_{0.4}$Na$_{0.6}$Fe$_2$As$_2$ crystals were always obtained once Al$_2$O$_3$ crucible is used to hold the melt but LaFeAsO crystals result when Ta crucible is used, even though we started with the same starting materials and used identical heat treatment process. The difference is illustrated in Figure\,\ref{LaNaFeAs-1}. For the growth of LaFeAsO single crystals, the starting materials of a homogeneous mixture of LaAs, (1/3Fe+1/3Fe$_2$O$_3$) or FeO, and NaAs with the molar ratio of 1:1:15 was loaded into a Ta tube.\cite{YanAPLGrowth} For the growth of La$_{0.4}$Na$_{0.6}$Fe$_2$As$_2$ single crystals, the same homogeneous mixture was loaded into an Al$_2$O$_3$ crucible, which was further sealed inside of a Ta tube under ~1/3 atmosphere of argon gas. The use of Ta tube in both cases is necessary to protect quartz tube from NaAs at high temperatures.

As shown in Figure\,\ref{LaNaFeAs-1}, the only difference between the above two growths is the use of an Al$_2$O$_3$ crucible which has a direct contact with the flux. The dramatic difference in the crystal composition  forces us to consider the possibility that the Al$_2$O$_3$ crucible absorbs oxygen introduced by iron oxide into the melt. The formation of NaAsO$_2$ has been proposed to account for the solubility of oxygen in an NaAs flux. NaAsO$_2$ has a low melting temperature of about 600$^o$C, which is similar to that of NaAs.\cite{YanFluxRequirement} After crystal growth, the Al$_2$O$_3$ crucible turns to be grey and reaction between the crucible and flux was always observed. Sometimes, the Al$_2$O$_3$ growth crucible cracks after growth. Residual flux was found both inside and outside of the growth crucible.  Unfortunately, we did not further investigate the reaction mechanism for the following two reasons:  (1) in a typical growth, about 0.10 gram Fe$_2$O$_3$ was used in a 3 gram mixture of charge and flux.  This can lead to the formation of 0.12 gram NaAsO$_2$.  With that little amount of NaAsO$_2$, it's difficult to find the reacting product in the ampoule;  and (2) sodium arsenate hydrate, which is poisonous and can lead to stomach pain, nausea, decreased blood pressure, headache, and possibly carcinogenic and teratogenic problems,  might form and stay on the surface of growth crucible when washing the crystals out of flux. So we tried to reduce any contact with the crucible and residual chemicals. Even without any direct observation of the reacting product, it is reasonable to believe that NaAsO$_2$ reacts with alumina crucible at high temperatures, considering the color change and cracking of the Al$_2$O$_3$ growth crucible, and various quaternary phases in Na-Al-As-O system. This reaction consumes oxygen in the melt and maintains an oxygen free environment. In contrast, the successful growth of LaFeAsO single crystals from NaAs flux in a Ta tube suggests that the melt cannot oxidize Ta tube at the growth conditions.

La$_{0.4}$Na$_{0.6}$Fe$_2$As$_2$  is a new ThCr$_2$Si$_2$-type ('122') compound with La$^{3+}$ and Na$^+$ ions occupying the same crystallographic site.  From a simple electron counting, varying the La/Na ratio in La$_{0.5-x}$Na$_{0.5+x}$Fe$_2$As$_2$ can tune the material to be either electron-doped (x$<$0) or hole-doped (x$>$0), which allows the study of electron-hole asymmetry without disturbing the FeAs conducting layer.  There are other 122 compounds with alkali metal, such as AFe$_2$As$_2$ (A=Alkali metal),\cite{KFe2As2,RbFe2As2,CsFe2As2} AE$_{1-x}$Na$_x$Fe$_2$As$_2$ (AE=alkaline earth ion).\cite{CaNaFe2As2,BaNaFe2As2,EuNaFe2As2} Crystal growth of 122 compounds with alkali ions is normally performed using a similar growth setup with an Al$_2$O$_3$ crucible, a Ta tube, and a quartz tube (see Fig 4(a)). The quartz tube protects the Ta tube from oxidization, and Ta tube prevents the evaporation loss of active alkali metals and reaction with quartz tube. At high temperatures, attack of Al$_2$O$_3$ crucible by alkali metals takes place leading to a variation of the crystal composition from the nominal one. But to the best of our knowledge, there is no report of oxygen contamination resulting from the reaction. For the growth of other 122 compounds without volatile alkali ions, little reaction was observed between flux and Al$_2$O$_3$ growth crucibles.\cite{BaFe2As2} The reaction observed when growing La$_{0.4}$Na$_{0.6}$Fe$_2$As$_2$ suggests if a low-melting compound containing oxygen forms, like NaAsO$_2$ in NaAs flux, oxide crucible might not be the suitable container for the growth of oxygen bearing compounds. However, this mechanism could be employed to eliminate oxygen introduced by starting materials to improve crystal quality.

\section{Summary}
In summary, we reviewed the flux growth of La$_5$Pb$_3$O and La$_{0.4}$Na$_{0.6}$Fe$_2$As$_2$ single crystals and illustrated how the reaction between flux and Al$_2$O$_3$ crucible controls oxygen content in the melt and the resulting crystals. For the case of La$_5$Pb$_3$O, Al$_2$O$_3$ crucible oxidizes La metal in the melt; this reaction forms a La$_2$O$_3$ shell which serves as the oxygen source to the melt and a passivating layer which protects the remaining La metal in the melt. For the case of La$_{0.4}$Na$_{0.6}$Fe$_2$As$_2$, the reaction between NaAsO$_2$ in the flux and Al$_2$O$_3$ crucible helps maintain an oxygen free environment to facilitate the growth of La$_{0.4}$Na$_{0.6}$Fe$_2$As$_2$. These two growth examples demonstrate that the reaction between flux and Al$_2$O$_3$ crucible can be employed to assist crystal growth, especially for growths during which the oxygen content in the high temperature melt should be controlled. This provides a new way to look for a proper flux for the growth of some special intermetallic compounds bearing oxygen, such as \emph{R}FeAsO superconductors.

\section{Acknowledgment}
JQY thanks P. C. Canfield, A. F. May, R. W. McCallum, and M. A. McGuire for helpful discussions and improving the manuscript. Work at ORNL was supported by the US Department of Energy, Office of Science, Basic Energy Science, Materials Sciences and Engineering Division. Part of the growth of La$_{0.4}$Na$_{0.6}$Fe$_2$As$_2$  was performed at Ames Laboratory.

\end{document}